\documentclass[%
 reprint,
superscriptaddress,
showpacs,
 amsmath,amssymb,
 aps,
]{revtex4-1}

\usepackage{graphicx}
\usepackage{color}
\usepackage{verbatim}
\usepackage{subfigure}

\DeclareMathOperator{\sn}{sn}
\DeclareMathOperator{\arcsn}{arcsn}

\definecolor{Red}{rgb}{1,0,0}

\begin{document}

\preprint{APS/123-QED}

\title{Symmetry-based analytical solutions to the $\chi^{(2)}$ nonlinear directional coupler} 

\author{David Barral} 
\email{Corresponding author: david.barral@c2n.upsaclay.fr}
\affiliation{Centre de Nanosciences et de Nanotechnologies C2N, CNRS, Universit\'e Paris-Saclay, 10 Boulevard Thomas Gobert, 91120 Palaiseau, France}
\author{Kamel Bencheikh}
\affiliation{Centre de Nanosciences et de Nanotechnologies C2N, CNRS, Universit\'e Paris-Saclay, 10 Boulevard Thomas Gobert, 91120 Palaiseau, France}
\author{Peter J. Olver}
\affiliation{{School of Mathematics, University of Minnesota, MN 55455 Minneapolis, U.S.A.}}
\author{Nadia Belabas}
\affiliation{Centre de Nanosciences et de Nanotechnologies C2N, CNRS, Universit\'e Paris-Saclay, 10 Boulevard Thomas Gobert, 91120 Palaiseau, France}
\author{Juan Ariel Levenson}
\affiliation{Centre de Nanosciences et de Nanotechnologies C2N, CNRS, Universit\'e Paris-Saclay, 10 Boulevard Thomas Gobert, 91120 Palaiseau, France}

\begin{abstract} In general the ubiquitous $\chi^{(2)}$ nonlinear directional coupler, where nonlinearity and evanescent coupling are intertwined, is nonintegrable. We rigorously demonstrate that matching excitation to the even or odd fundamental supermodes yields dynamical analytical solutions for any phase matching in a symmetric coupler. We analyze second harmonic generation and optical parametric amplification regimes and study the influence of fundamental fields parity and power on the operation of the device. These fundamental solutions are useful to develop applications in classical and quantum fields such as all-optical modulation of light and quantum-states engineering.
\end{abstract}

\date{February 04, 2019}
\maketitle 

{The nonlinear directional coupler (NDC) is a core device in integrated optics. Its potential was first demonstrated in $\chi^{(3)}$ materials as an all-optical switch \cite{Jensen1982, Maier1983}. This and other interesting functionalities were later displayed in the $\chi^{(2)}$ NDC through cascaded second-order effects \cite{Assanto1993, Schiek1994, Schieck1996, Schiek1999, Hempelmann2002}. In the last years the $\chi^{(2)}$ NDC has found a flourishing field of application: quantum optics \cite{Perina2000}. Its key strengths in quantum information processing as a source of entangled photons and entangled field quadratures have been demonstrated and are still actively explored \cite{Herec2003, Kruse2013, Kruse2015, Setzpfandt2016, Barral2017, Barral2018}. In general the $\chi^{(2)}$ NDC is a nonintegrable system and only stationary solutions --solitons-- are available \cite{Mak1997, Bang1997, Mak1998}. Even in this case, general solutions are only obtained numerically or in a semianalytical form \cite{Mak1998b}. The dynamical solutions of the $\chi^{(2)}$ NDC have nonetheless a broad range of applications in the classical and quantum regimes \cite{Assanto1993, Schiek1994, Schieck1996, Schiek1999, Perina2000, Herec2003, Kruse2013, Kruse2015, Setzpfandt2016, Barral2017, Barral2018}. Two limiting cases only have up to now been identified as integrables, i.e. with analytical dynamical solutions: (i) The propagation equations can be reduced to those related to the simpler $\chi^{(3)}$ NDC when the phase mismatch between the fundamental and second harmonic waves propagating in the device is large, which corresponds to a regime of lower efficiency \cite{Bang1997}. (ii) The undepleted harmonic-field approximation in spontaneous parametric downconversion linearizes the propagation equations \cite{Perina2000}. }

Analytical solutions are universally preferred since they can be used to contemplate new applications and engineer the propagation of both classical and quantum light in these devices. In this paper, we rigorously retrieve analytical solutions for the $\chi^{(2)}$ NDC for --any--  phase matching under specific symmetry conditions: pumping in the even or odd fundamental supermode. We show indeed that the propagation equations are analogous to those related to a single $\chi^{(2)}$ nonlinear waveguide with imperfect phase matching. We show that in the NDC case the effective coupling plays the role of the wavevector phase mismatch in the emblematic single waveguide \cite{Armstrong1962}. We can thus analyze second harmonic generation (SHG) and optical parametric amplification (OPA) in this configuration, shedding light on the influence of total power and fundamental-modes phases on the operation of the device, towards higher efficiency and quantum applications.

\begin{figure}[t]
\centering
\includegraphics[width=0.48\textwidth]{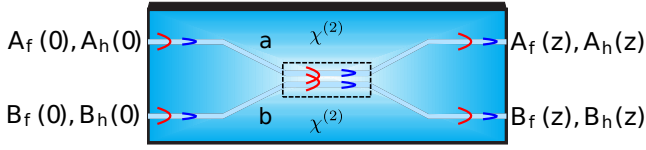}
\vspace {0cm}\,
\hspace{0cm}\caption{\label{F1}\small{(Color online) Sketch of the nonlinear directional coupler $\chi^{(2)}$-NDC made of two identical waveguides $\it{a}$ and $\it{b}$ with second-order susceptibilities $\chi^{(2)}$. The dashed box indicates the nonlinear and coupling region. Input fundamental fields produce second harmonic fields through SHG or input harmonic fields amplify injected fundamental seeds through OPA. In red, the fundamental waves, evanescently coupled (f). In blue, and more confined, the non-interacting second harmonic waves (h).}}
\end {figure}

The $\chi^{(2)}$ NDC, sketched in Figure \ref{F1} (dashed box), is made of two identical nonlinear $\chi^{(2)}$ waveguides. In each waveguide, an input fundamental field at frequency $\omega_{f}$ is up-converted into a second-harmonic field at frequency $\omega_{h}$ (SHG), or a weak input fundamental field is amplified with the help of a strong second-harmonic field (degenerate OPA). For the sake of simplicity, we consider all fields in the same polarization mode. In the coupling region, the energy of the fundamental modes propagating in each waveguide is exchanged between the coupled waveguides through evanescent waves, whereas the interplay of the generated, or injected, second harmonic waves is negligible for the considered propagation lengths due to their high confinement into the waveguides. Both physical processes, evanescent coupling and nonlinear generation, are described by the following system of equations \cite{Assanto1993}
\begin{align} \label{QS1}
\frac{d {A}_{f}}{d z}=&\, i C {B}_{f} +2 i g {A}_{h} {A}_{f}^{*}\,e^{i \Delta\beta z}, \quad \frac{d {A}_{h}}{d z}= i g {A}_{f}^{2}\,e^{-i \Delta\beta z}, \nonumber \\
\frac{d {B}_{f}}{d z}=&\, i C {A}_{f} +2 i g {B}_{h} {B}_{f}^{*}\,e^{i \Delta\beta z}, \quad \frac{d {B}_{h}}{d z}= i g {B}_{f}^{2}\,e^{-i \Delta\beta z},
\end{align}
where $A$ and $B$ are the slowly varying amplitudes of fundamental ({\it f}) and second harmonic ({\it h}) fields corresponding to the upper (a) and lower (b) waveguides, respectively, $g$ is the nonlinear constant proportional to $\chi^{(2)}$ and the spatial overlap of the fundamental and harmonic fields in each waveguide, $C$ the linear coupling constant, $\Delta\beta\equiv\beta(\omega_{h})-2\beta(\omega_{f})$ the wavevector phase mismatch with $\beta(\omega)$ the propagation constant at frequency $\omega$, and $z$ is the coordinate along the direction of propagation. $C$ and $g$ are taken as real without loss of generality. We consider $C=8 \times 10^{-2}$ \,mm$^{-1}$, $g=25 \times 10^{-4}$ \,mm$^{-1}$ mW$^{-1/2}$ and lengths of few centimeters in the simulations we show below. These are state-of-the-art values in periodically poled lithium niobate waveguides  \cite{Alibart2016}. The input powers used in the simulations are of the order of those in \cite{Schiek1999}.

In order to solve the set of Equations (\ref{QS1}), we use dimensionless amplitudes and phases related to the complex amplitudes through 
\begin{align} 
u_{f}(z)&=\frac{\vert A_{f}(z)\vert}{\sqrt{P}},  &  &v_{f}(z)=\frac{\vert B_{f}(z)\vert}{\sqrt{P}}, \nonumber \\
\theta_{f}(z)&=\arg\{A_{f}(z)\}, &  &\phi_{f}(z)=\arg\{B_{f}(z)\}, \nonumber \\
u_{h}(z)&=\sqrt{\frac{2}{P}}\vert A_{h}(z)\vert,  &  &v_{h}(z)=\sqrt{\frac{2}{P}}\vert B_{h}(z)\vert, \nonumber \\
\theta_{h}(z)&=\arg\{A_{h}(z)\}, &  &\phi_{h}(z)=\arg\{B_{h}(z)\}, \nonumber
\end{align}
with $P=\vert A_{f} \vert^{2} + \vert B_{f} \vert^{2}+2\vert A_{h} \vert^{2} + 2\vert B_{h} \vert^{2}$ the total input power. We also introduce a normalized propagation coordinate $\zeta\equiv\sqrt{2 P} g z$, which is defined only in the nonlinear and coupling region (Figure 1, dashed box). {Applying this change of variables into Equations (\ref{QS1}), we obtain for the modes propagating in waveguide $\it{a}$
\begin{align}\label{us1}
\frac{d {u}_{f}}{d \zeta}=&  -\kappa \, {v}_{f} \sin(\delta_{f}) - u_{f} u_{h} \sin(\Delta\theta),   \\ \label{us2}
\frac{d {\theta}_{f}}{d \zeta}=& \,\kappa \frac{{v}_{f}}{u_{f}} \cos(\delta_{f}) + u_{h} \cos(\Delta\theta), \\ \label{us3}
\frac{d {u}_{h}}{d \zeta}=& \,u_{f}^2 \sin(\Delta\theta), \\ \label{us4}
\frac{d {\theta}_{h}}{d \zeta}=&\, \frac{u_{f}^2}{u_{h}} \cos(\Delta\theta), 
\end{align}
and for the modes propagating in waveguide $\it{b}$
\begin{align}\label{ys1}
\frac{d {v}_{f}}{d \zeta}=&  \kappa \, {u}_{f} \sin(\delta_{f}) - v_{f} v_{h} \sin(\Delta\phi),   \\ \label{ys2}
\frac{d {\phi}_{f}}{d \zeta}=& \,\kappa \frac{{u}_{f}}{v_{f}} \cos(\delta_{f}) + v_{h} \cos(\Delta\phi), \\ \label{ys3}
\frac{d {v}_{h}}{d \zeta}=& \,v_{f}^2 \sin(\Delta\phi), \\ \label{ys4}
\frac{d {\phi}_{h}}{d \zeta}=&\, \frac{v_{f}^2}{v_{h}} \cos(\Delta\phi).
\end{align}
The three governing parameters of the system are the effective coupling $\kappa\equiv C/(\sqrt{2P} g)$, the fundamental fields phase difference $\delta_{f}\equiv\phi_{f}-\theta_{f}$, and the nonlinear phase mismatchs $\Delta\theta \equiv \theta_{h}-2\theta_{f}+ \Delta S \zeta$ and $\Delta\phi \equiv \phi_{h}-2\phi_{f}+ \Delta S \zeta$, where $\Delta S=\Delta\beta/(\sqrt{2P} g)$ is an effective wavevector phase mismatch. Remarkably, the nonlinear phase mismatch drives the nonlinear optical processes whereas the effective coupling indicates which effect is stronger, either the evanescent coupling or the nonlinear interaction. Additionally, there are two dynamical invariants, the energy and momentum of the total system given respectively by 
\begin{align}  \label{DynInv1}
&u_{f}^{2} + v_{f}^{2} + u_{h}^{2} + v_{h}^{2} =1,  \\   \label{DynInv2}
&u_{h} u_{f}^{2} \cos (\Delta\theta) + v_{h} v_{f}^{2} \cos (\Delta\phi) + 2 \kappa u_{f} v_{f} \cos(\delta_{f}) = \Gamma,
\end{align}
where $\Gamma$ is a constant given by the initial conditions \cite{Note1}. 

The systems of Equations (\ref{us1}-\ref{us4}) and (\ref{ys1}-\ref{ys4}) are not integrable in general \cite{Bang1997}. The key to our analytical solution is to take advantage of the fact that the full system of Equations (\ref{us1}-\ref{ys4}) is invariant under the following set of transformations $F(u_{f}, \theta_{f}, u_{h}, \theta_{h}, v_{f}, \phi_{f}, v_{h}, \phi_{h})$:
\begin{align}
u_{f}&\leftrightarrow v_{f}, &u_{h}\leftrightarrow v_{h},  \nonumber \\
\phi_{f}&\leftrightarrow \theta_{f}+n\pi, &\Delta\theta \leftrightarrow \Delta\phi,  \label{Symmetry}
\end{align}
with $n=0,1$. The two last transformations can be combined to obtain $\phi_{h} \leftrightarrow \theta_{h}$. In general this set of transformations modifies the initial conditions of the problem, thus losing the symmetry and the dynamical invariance. Nonetheless, we crucially notice that for symmetric initial conditions
\begin{align}
u_{f}(0)&= v_{f}(0),  &u_{h}(0)= v_{h}(0), \nonumber \\
\phi_{f}(0)&=\theta_{f}(0)+n\pi, &\phi_{h}(0)= \theta_{h}(0), \label{InConditions}
\end{align}
the symmetry relations between the fields amplitudes and phases persist along propagation, protected by the invariance of the Equations (\ref{us1}-\ref{ys4}), so that at all $z$ 
\begin{align}
u_{f}&= v_{f},  &u_{h}= v_{h}, \nonumber \\
\phi_{f} &=\theta_{f}+n\pi, &\phi_{h}= \theta_{h}. \label{Conditions}
\end{align}
These relations were mentioned along the analysis of the stationary solutions to Equations (\ref{QS1}) \cite{Bang1997}. However, the connection between the initial conditions Equations (\ref{InConditions}) and the solutions Equations (\ref{Conditions}) was missing. We proceed here to give rigorous justification to Equations (\ref{Conditions}). Let us rewrite Equations (\ref{us1}-\ref{us4}) and (\ref{ys1}-\ref{ys4}) as a single vector equation
\begin{equation} \label{sym}
 \frac{d\mathbf{x}}{d\zeta}=h(\mathbf{x}),
\end{equation} 
with $\mathbf{x}=(u_{f}, \theta_{f}, u_{h}, \theta_{h}, v_{f}, \phi_{f}, v_{h}, \phi_{h})^{T}$. Let $F$ be the locally defined invertible differentiable map which is given by Equations (\ref{Symmetry}). Then, by the chain rule, $\mathbf{y}(\zeta)=F(\,\mathbf{x}(\zeta))$ solves the system of ordinary differential equations 
\begin{equation}\nonumber
d\mathbf{y}/d\zeta=H(\mathbf{y})=\nabla F (F^{-1}(\mathbf{y})) h(F^{-1}(\mathbf{y})),
\end{equation}
where $\nabla F(\mathbf{x})$ denotes the Jacobian matrix of $F$ at $\mathbf{x}$. Now suppose $H=h$ on their common domain of definition, meaning that the map $F$ defines a symmetry of Equation (\ref{sym}). Furthermore, suppose $F$ is also a symmetry of the initial conditions $\mathbf{x}(0)=\mathbf{x}_{0}$ such that $F(\mathbf{x}_{0})=\mathbf{x}_{0}$. Then $\mathbf{x}({\zeta})$ and $\mathbf{y}(\zeta)$ both solve the same initial value problem. Hence, since the system is smooth (indeed analytic), by uniqueness of solutions to the initial value problem, they must be the same, meaning that $F$ is also a symmetry of the solution
\begin{equation} \label{proof}
\mathbf{x}(\zeta)=F(\,\mathbf{x}(\zeta)),
\end{equation}
which proves Equations (\ref{Conditions}). This proof is indeed general: any system with smooth evolution and invariant under an invertible and differentiable transformation $F$ has solutions that retain this invariance provided the initial conditions are also $F$-invariant. This result is related to century-old questions concerning the impact of symmetries on physical systems, formulated by P. Curie and S. Lie \cite{Ismael1997}.


The symmetry of the solutions Equation (\ref{proof}) thus simplifies the system of Equations  (\ref{us1}-\ref{ys4}) into
\begin{align}\label{vs1}
\frac{d {u}_{f}}{d \zeta}=& - u_{f} u_{h} \sin(\Delta\theta),   \\ \label{vs2}
\frac{d {u}_{h}}{d \zeta}=& \,u_{f}^2 \sin(\Delta\theta), \\ \label{vs3}
\frac{d {\Delta\theta}}{d \zeta}=& \,(\frac{u_{f}^2}{u_{h}}- 2u_{h}) \cos(\Delta\theta) - (-1)^{n} 2 \kappa,
\end{align}
and the dynamical invariants Equations (\ref{DynInv1}-\ref{DynInv2}) into
\begin{align} \label{vs4}
u_{f}^{2} + u&_{h}^{2} =1/2, \,\, \,\qquad v_{f}^{2} + v_{h}^{2} =1/2,\\ \label{vs5}
(1- 2u&_{h}^{2}) (u_{h} \cos (\Delta\theta) + (-1)^{n} \kappa) = \Gamma. 
\end{align}
Remarkably, these equations are analogous to those related to the nonlinear interaction of two waves with imperfect phase matching $\Delta\beta$ in a bulk crystal or single waveguide \cite{Armstrong1962}. In our case, the effective coupling $2\kappa$ plays the role of $\Delta \beta$ in the crystal or single waveguide. The reduced Equations (\ref{vs1}-\ref{vs3}) are fulfilled only when harmonic and fundamental input powers are set equal in each waveguide, $u_{f}^{2}(0)=v_{f}^{2}(0)$ and $u_{h}^{2}(0)=v_{h}^{2}(0)$, harmonic fields in phase, $\theta_{h}(0)=\phi_{h}(0)$, and fundamental fields either in phase, $\theta_{f}(0)=\phi_{f}(0)$, or $\pi$-dephased, $\theta_{f}(0)=\phi_{f}(0) + \pi$. This leads to a  reasonable set of initial conditions for the $\chi^{(2)}$ NDC as these conditions correspond to the excitation of the even or odd fundamental eigenmodes of the linear directional coupler, so-called supermodes \cite{Yariv1988}. Outstandingly, the $\chi^{(2)}$ NDC is a versatile source of quantum entanglement under these conditions \cite{Herec2003, Barral2017, Barral2018}. 

Equations (\ref{vs1}-\ref{vs3}) have analytical solutions in terms of Jacobi elliptic functions \cite{Armstrong1962}. We analyze thoroughly these solutions in the SHG and OPA regimes. From Equations (\ref{vs2}) and (\ref{vs5}), we get
\begin{equation}\label{Sol1}
\zeta=\pm\frac{1}{2} \int_{u_{h}^{2}(0)}^{u_{h}^{2}(\zeta)}  \, \frac{d (u_{h}^{2})}{\sqrt{u_{h}^{2}(\frac{1}{2} - u_{h}^{2})^{2}-[\frac{\Gamma}{2} -(-1)^{n} \kappa\, (\frac{1}{2} - u_{h}^{2})]^2}}.
\end{equation}
The expression in the square root has three roots $u_{h,3}^{2}>u_{h,2}^{2}>u_{h,1}^{2}\geq0$. By using the function $\it{y}$ and the parameter $\gamma$, defined respectively as $y^{2}=(u_{h}^{2} - u_{h,1}^{2})/(u_{h,2}^{2} - u_{h,1}^{2})$ and $\gamma^{2}=(u_{h,2}^{2} - u_{h,1}^{2})/(u_{h,3}^{2} - u_{h,1}^{2})$, we can rewrite Equation (\ref{Sol1}) as
\begin{equation}\nonumber
\zeta=\frac{\pm 1}{2 \sqrt{u_{h,3}^{2} - u_{h,1}^{2}}} \int_{y(0)}^{y(\zeta)}  \, \frac{d y}{\sqrt{(1-y^{2})(1-\gamma^{2} y^{2})}}.
\end{equation}
$\it{y}$ is the Jacobi elliptic function of $\zeta$. The normalized harmonic power is thus given by
\begin{equation}\label{Sol3}
u_{h}^{2}=u_{h,1}^{2}+(u_{h,2}^{2}-u_{h,1}^{2}) \,\sn^{2}(\sqrt{u_{h,3}^{2}-u_{h,1}^{2}}(\zeta + \zeta_{0}),\gamma),
\end{equation}
where $\sn$ stands for the Jacobi elliptic sine. 
$\zeta_{0}$ is determined by the initial condition $u_{h}^{2}(0)$ and the parameter $\gamma$, and it is given by
\begin{equation}\nonumber
\zeta_{0}=\frac{1}{\sqrt{u_{h,3}^{2}-u_{h,1}^{2}}} \, \arcsn(\sqrt{\frac{u_{h}^{2}(0)-u_{h,1}^{2}}{u_{h,2}^{2}-u_{h,1}^{2}}}, \gamma),
\end{equation}
where $\arcsn$ stands for the inverse Jacobi elliptic sine. The period of oscillations in the harmonic powers is thus
\begin{equation}\label{PO}
L=\frac{2 K(\gamma)}{\sqrt{u_{h,3}^{2}-u_{h,1}^{2}}},
\end{equation}
with $\it{K}$ the complete elliptic integral of first kind. The individual phases $\theta_{f,h}$ and the nonlinear phase mismatch $\Delta\theta$ can be straightforwardly obtained from Equations (\ref{us2}), (\ref{us4}) and  (\ref{Sol3}), and the invariants given by Equations (\ref{vs4}) and (\ref{vs5}). 

\begin{figure}[t]
\centering
\includegraphics[width=0.47\textwidth]{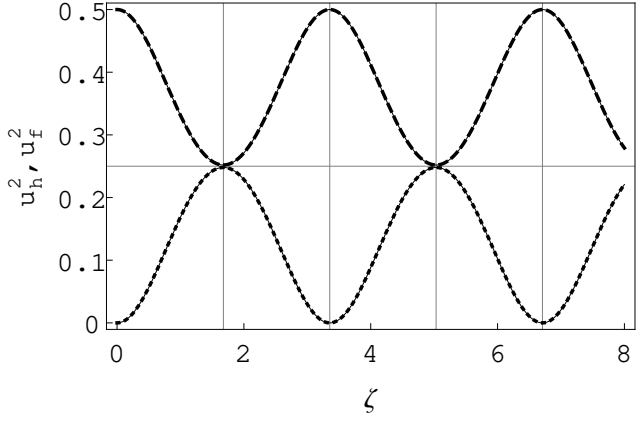}
\vspace {-0.25cm}\,
\hspace{0cm}\caption{\label{F2}\small{Fundamental (upper curve) and harmonic (lower curve) fields power propagation in the SHG regime. Dimensionless fundamental powers $u_{f}^{2}$: analytical (dash) and numerical (solid). Dimensionless second harmonic powers $u_{h}^{2}$: analytical (dot) and numerical (solid). $\kappa=0.51$. The vertical lines show the effective coupling coherence length $L/2$, with $L=3.35$ the oscillation period analytically calculated. $\zeta$ is the normalized propagation coordinate. $\zeta=1$ stands for $z\equiv(\sqrt{2 P} g)^{-1}=6.3$ mm.}}

\end {figure}
\begin{figure}[t]
\centering
\includegraphics[width=0.47\textwidth]{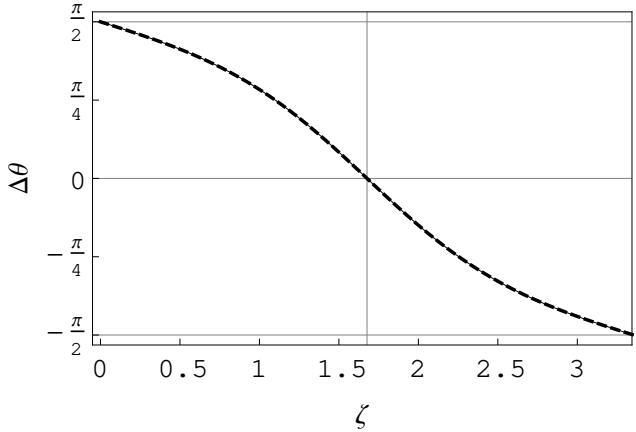}
\vspace {-0.25cm}\,
\hspace{0cm}\caption{\label{F3}\small{Nonlinear phase mismatch evolution along propagation in one oscillation period $L$ in the SHG regime. Analytical (dash) and numerical (solid). $\kappa=0.51$ and $n=0$. The vertical line shows the effective coupling coherence length $L/2$, with $L=3.35$ the oscillation period analytically calculated. $\zeta$ is the normalized propagation coordinate. $\zeta=1$ stands for $z\equiv(\sqrt{2 P} g)^{-1}=6.3$ mm.}}
\end {figure}

Now we show the solutions for two specific cases of SHG and OPA. We consider perfect wavevector phase matching $\Delta\beta=0$ in both situations for the sake of simplicity. For SHG $u_{h}^{2}(0)=0$ and $\Gamma=(-1)^{n}\kappa$, so that the roots from the expression in the square root of Equation (\ref{Sol1}) are solutions of
\begin{equation}\nonumber
u_{h}^{6} - (1+\kappa^{2}) u_{h}^{4} + \frac{u_{h}^{2}}{4}=0,
\end{equation}
which read
\begin{equation}\label{Sol5}
u_{h,1}^{2}=0, \quad u_{h, 2(3)}^{2}=\frac{1+\kappa^{2}\pm\kappa\sqrt{2+\kappa^{2}}}{2}.
\end{equation}
Notably, these solutions depend only on the strength of the effective coupling $\kappa$ and not on the supermode parity, i.e. not on $n$. This point is clarified in the analysis of the nonlinear phase mismatch evolution below.

Figure \ref{F2} displays the dimensionless analytically (Eqs. (\ref{Sol3}) and (\ref{Sol5})) and numerically (Eqs. (\ref{us1}-\ref{ys4})) calculated powers for each mode in waveguide $\it{a}$ (or equally $\it{b}$) along the propagation in the SHG regime. We have set $\kappa=0.51$, equivalent to $P=2\,W$ for our realistic values in Lithium Niobate. A strong fundamental field depletion and a periodic switch from fundamental-to-harmonic conversion to harmonic-to-fundamental conversion are observed. $L=3.35$ is the period of oscillation analytically calculated through Equation (\ref{PO}). $L/2$ is the effective coupling coherence length defined in analogy with the wave-vector coherence length. The connection of the observed periodic behaviour and a coupling-based nonlinear phase mismatch has been proposed recently through the analysis of numerical simulations \cite{Barral2017, Barral2018}. To clarify the origin of these periodic oscillations, we calculate the evolution of phases along propagation. The individual phases are given by 
\begin{align}\nonumber
&\theta_{h}(\zeta)=\theta_{h}(0)+(-1)^{n} \kappa \,\zeta,\\  \label{Phases}
&\theta_{f}(\zeta)=\theta_{f}(0)+\frac{(-1)^{n} \kappa}{u_{h,3}} \,\Pi(2u_{h,2}^{2},\Phi(u_{h,3}\zeta, \gamma),\gamma),
\end{align}
where $\Pi$ is the elliptic integral of the third kind and $\Phi$ the amplitude of Jacobi elliptic functions. Figure \ref{F3} shows analytically and numerically calculated evolution of the nonlinear phase mismatch $\Delta\theta(\zeta)$ in a period of oscillation $L$. We set $\kappa$ as above, $\theta_{f}(0)=0$ and $\theta_{h}(0)=\pi/2$ due to the well-known SHG phase jump \cite{Armstrong1962}. The phase mismatch evolves from $\pi/2$ down to $-\pi/2$ in an oscillation period $L$ when the parity is set as $n=0$ (Figure {\ref{F3}}). A symmetric evolution curve from $\pi/2$ up to $3\pi/2$ is obtained for $n=1$ (not shown). Since the evolution of $\sin(\Delta\theta)$, and thus of the $u_f$ and $u_h$ solutions in Equations (\ref{vs1})-(\ref{vs2}), is the same in both cases, SHG is independent of the input supermode parity. Equations (\ref{Phases}) show that the linear coupling of the fundamental modes $\kappa$ produces a nonlinear phase mismatch which cyclically destroys the wavevector phase matching initially fulfilled, driving two successive nonlinear optical processes, upconversion in the first effective coupling coherence length followed by downconversion in the second coupling length. Note that numerical and analytical solutions perfectly match \cite{Note2}. 

For OPA with a set of input phases such that $\Delta\theta(0)=0$, $\Gamma=u_{h}(0)-2u_{h}^{3}(0)+(-1)^{n}\kappa\,(1-2u_{h}^{2}(0))$ is preserved along propagation, and the roots of the expression in the square root of Equation (\ref{Sol1}) are given by solving the expression
\begin{equation}\nonumber
u_{h}^{2}(\frac{1}{2} - u_{h}^{2})^{2}-[\frac{u_{h}(0)}{2}-u_{h}^{3}(0)+(-1)^{n}\kappa\,( u_{h}^{2} -u_{h}^{2}(0) )]^2=0,
\end{equation}
with general solutions
\begin{widetext}
\begin{equation}\label{Sol6}
u_{h,1}^{2}=u_{h}^{2}(0),\quad
u_{h, 2(3)}^{2}=\frac{1}{2}(1-u_{h}^{2}(0)+\kappa^{2}\mp\sqrt{2u_{h}^{2}(0)(1-3\kappa^{2}-\frac{3}{2}u_{h}^{2}(0))+4 (-1)^{n}\kappa\,u_{h}^{2}(0)(1-2u_{h}^{2}(0))+\kappa^{2}(2+\kappa^{2})}).
\end{equation}
\end{widetext}
These solutions also include SHG when $u_{h}^{2}(0)=0$. Note that Equations (\ref{Sol6}) have to be suitably ordered in order to be used in Equation (\ref{Sol3}). In contrast to SHG, Equations (\ref{Sol6}) depend in this case on the input harmonic power $u_{h}^{2}(0)$, the effective coupling $\kappa$ and the parity of the input fundamental supermode via the parameter $\it{n}$. We show below the dependence of the solutions on parity and input harmonic power.

Figure \ref{F4} top displays the dimensionless analytically (Eqs. (\ref{Sol3}) and (\ref{Sol6})) and numerically (Eqs. (\ref{us1}-\ref{ys4})) calculated fundamental powers in waveguide $\it{a}$ (or equally $\it{b}$) along the propagation in a specific case of OPA. We have set $u_{h}^{2}(0)=0.499$, $\kappa=0.92$ ($P= 600\,mW$) and $n=0$ (top figure, black dash for analytical, black solid for numerical) and $n=1$ (top figure, gray dash for analytical, gray solid for numerical). The harmonic fields are not shown since they remain almost undepleted for this set of parameters. Note that the power scale (ordinate axis) has been expanded by a factor of $10^{3}$. Numerical and analytical solutions perfectly match again. In contrast with SHG, OPA depends on the parity of the input fundamental supermodes. The system periodically switches from harmonic-to-fundamental conversion to fundamental-to-harmonic conversion for even input parity, whereas it switches from fundamental-to-harmonic conversion to harmonic-to-fundamental conversion for odd input parity. Unlike in SHG, the nonlinear phase mismatch $\Delta\theta$ evolves in OPA from the initial value $\Delta\theta(0)=0$ to negative  (n=0) or positive (n=1) values (not shown). This modifies the evolution of the amplitudes  $u_{f,h}$ through the sign of $\sin(\Delta\theta)$ in Equations (\ref{vs1})-(\ref{vs2}). The period of oscillation is also modified by the input parity with $L_\text{even}=5.19$ and $L_\text{odd}=5.27$. At $L_\text{odd}/2$ the fundamental odd mode is reduced by approximately a factor $7.5$, whereas at $L_\text{even}/2$ the fundamental even mode is amplified by the same factor. Figure \ref{F4} bottom displays the ratio between even and odd fundamental fields power $u_{f,e}^{2}/u_{f,o}^{2}$ along propagation. Notably, a ratio higher than 50 is obtained at the odd effective coupling coherence length $L_\text{odd}/2$. 

\begin{figure}[t]
  \centering
    \subfigure{\includegraphics[width=0.47\textwidth]{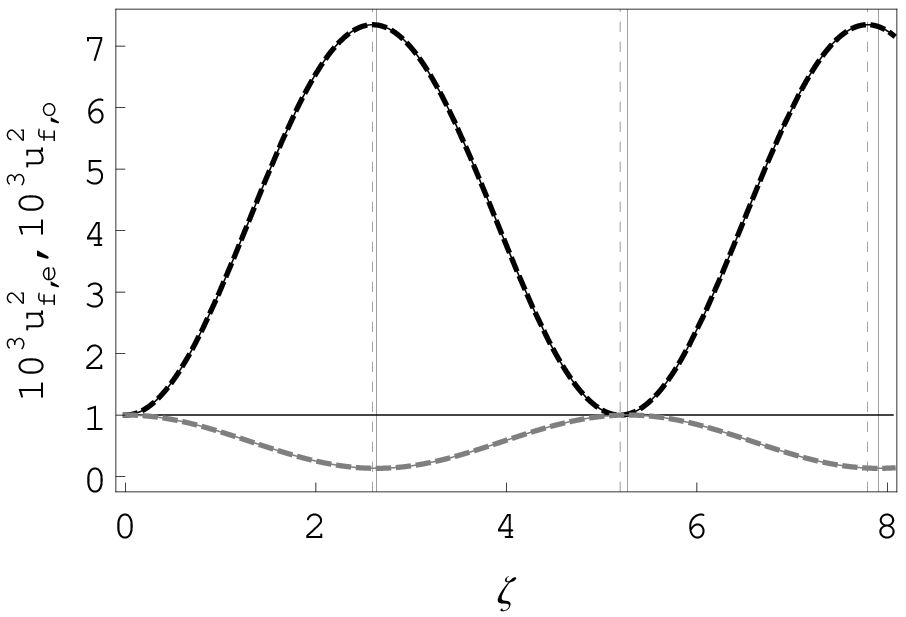}}
\vspace {0cm}\,
    \subfigure{\includegraphics[width=0.47\textwidth]{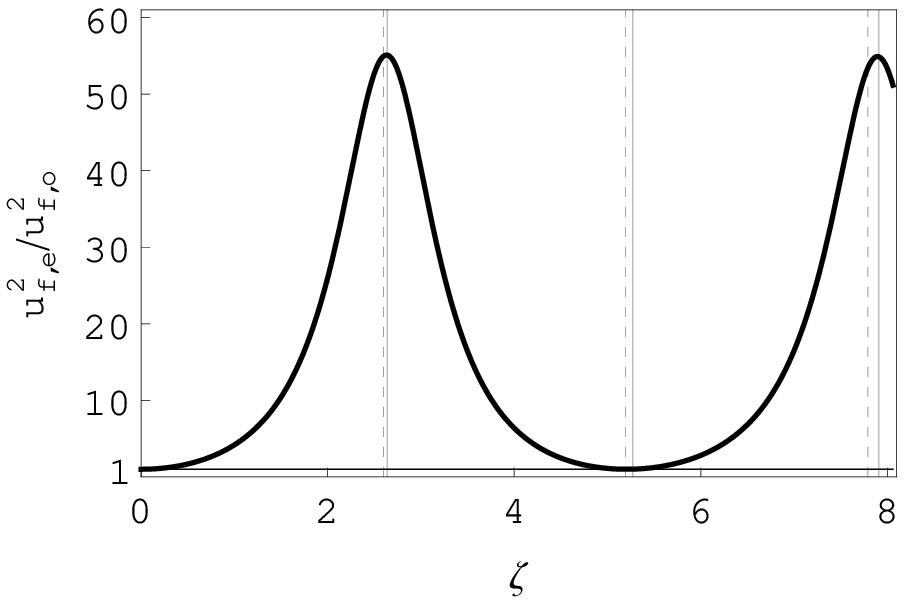}}
\vspace {0cm}\,
\hspace{0cm}\caption{\label{F4}\small{Top figure: fundamental fields power propagation in the OPA regime for an even (black) and odd (gray) input fundamental supermode. Even dimensionless fundamental power $u_{f,e}^{2}$: analytical (dash, black) and numerical (solid, black). Odd dimensionless fundamental power $u_{f,o}^{2}$: analytical (dash, gray) and numerical (solid, gray). Bottom figure: ratio between even and odd fundamental fields power $u_{f,e}^{2}/u_{f,o}^{2}$ along propagation. $\kappa=0.92$. The vertical lines show the even and odd effective coupling coherence lengths $L_{even}/2$ (dash) and $L_{odd}/2$ (solid), with $L_{even}=5.19$ and $L_{odd}=5.27$ analytically calculated. $\zeta$ is the normalized propagation coordinate. $\zeta=1$ stands for $z\equiv(\sqrt{2 P} g)^{-1}=11.5$ mm.}}
\end {figure}

\begin{figure}[t]
\centering
\includegraphics[width=0.47\textwidth]{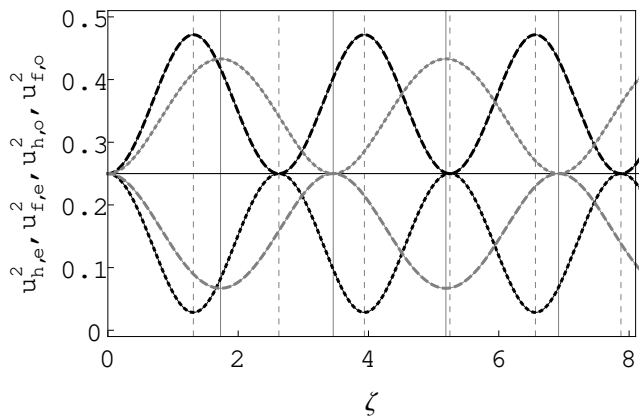}
\vspace {-0.25cm}\,
\hspace{0cm}\caption{\label{F5}\small{Fundamental (dash) and harmonic (dot) fields power propagation in the OPA regime for an even (black) and odd (gray) input fundamental supermode. Even dimensionless fundamental and harmonic power $u_{f(h),e}^{2}$: analytical (black dash (dot)) and numerical (black solid). Odd dimensionless fundamental and harmonic power $u_{f(h),o}^{2}$: analytical (gray dash (dot)) and numerical (gray solid). $\kappa=0.92$. The vertical lines show the even and odd effective coupling coherence lengths $L_{even}/2$ (dash) and $L_{odd}/2$ (solid), with $L_{even}=2.62$ and $L_{odd}=3.46$ analytically calculated. $\zeta$ is the normalized propagation coordinate. $\zeta=1$ stands for $z\equiv(\sqrt{2 P} g)^{-1}=11.5$ mm.}}
\end {figure}

Figure \ref{F5} displays the dimensionless analytically (Eqs. (\ref{Sol3}) and (\ref{Sol6})) and numerically (Eqs. (\ref{us1}-\ref{ys4})) calculated fundamental and harmonic powers in waveguide $\it{a}$ (or equally $\it{b}$) along propagation in the OPA regime for equal injection of fundamental and harmonic fields, i.e. $u_{h}^{2}(0)=0.25$. For the sake of comparison, the effective coupling is set as above ($\kappa=0.92$). We show the evolution of the fields produced by injection of even (black) or odd (gray) fundamental supermodes at the input. Fundamental and harmonic fields are in dash and dot, respectively. Strong harmonic fields depletion and fundamental fields amplification are achieved for even (n=0) input. Lower fundamental fields depletion and harmonic fields amplification are obtained for odd (n=1) input. Shorter periods of oscillation and larger even-odd oscillation period shifts are observed in comparison with those in Figure \ref{F4}. The even configuration allows to switch from harmonic undepletion to a large amount of depletion at $L_{even}/2$ by either injection of no, or very small, fundamental seed as in Figure~\ref{F4} or a substantial fundamental seed as in Figure~\ref{F5}. We have also found that the higher the total input power $P$, the larger the harmonic fields depletion (not shown). In contrast, the odd configuration yields the converse effect: the harmonic fields are amplified when substantial fundamental seeds are injected. Hence, two mechanisms, parity and power of the fundamental supermode, can be used as modulation parameters for a $\chi^{(2)}$ NDC all-optical switch. The analytical solutions enable prediction of the amplitude and period of oscillation of the optical fields along propagation through Equations (\ref{Sol3}) and (\ref{PO}), respectively. It is then possible to fix appropriately the initial conditions for the desired operating mode, even in the quantum regime \cite{Barral2017, Barral2018}.

In conclusion, we have studied the $\chi^{(2)}$ NDC and rigorously demonstrated that matching excitation to the even or odd fundamental supermodes yields dynamical analytical solutions for any phase matching. The propagation equations are analogous to those related to a single $\chi^{(2)}$ nonlinear waveguide with imperfect phase matching, but in the NDC we show that the effective coupling plays the role of the wavevector phase mismatch. We have reviewed the SHG and OPA regimes and studied the influence of fundamental fields parity and power on the operation of the device. We have investigated the possible application of this device as an all-optical switch. This study completes the analysis carried out in \cite{Barral2017, Barral2018}, where the versatility of this device as a resource for quantum information processing was shown. Finally, we want to stress that our analysis can open new avenues in the study of general coupled $\chi^{(2)}$ nonlinear systems, such as arrays of nonlinear waveguides in optics and Fermi resonance interface modes in solid state physics \cite{Setzpfandt2010, Agranovich2008}. The use of symmetries can indeed help to simplify these systems and obtain analytical solutions to understand their dynamics better.


{\it Acknowledgements.} We thank K. Belabas for useful discussions. 
This work was supported by the Agence Nationale de la Recherche through the INQCA project (grant agreement number PN-II-ID-JRP-RQ-FR-2014-0013), the Paris \^Ile-de-France region in the framework of DIM SIRTEQ through the project ENCORE, and the Investissements d'Avenir program (Labex NanoSaclay, reference ANR-10-LABX-0035).

\end{document}